\documentclass[aps,pra,11pt]{revtex4-2}

\usepackage{amsmath,amssymb,bm,caption,setspace,xspace,fancyvrb,scrextend,siunitx,euscript,color,xcolor,ulem}
\usepackage{enumerate,braket,float,wrapfig,booktabs,textcomp,comment}
\usepackage[pdftex]{graphicx}
\usepackage[displaymath]{lineno}
\usepackage[colorlinks]{hyperref}
\usepackage{notes2bib}

\newcommand{\look}{\marginpar{\bf{look}}}
\newcommand{\uu}[1]{\ensuremath{\, \mathrm{#1}}}
\newcommand{\chem}[1]{\ensuremath{\mathrm{#1}}}

\newcommand{\ra}{\ensuremath{\rightarrow}\xspace}
\let\v\bm
\def\sR{{\ensuremath{\mathcal R}}}
\def\sE{{\ensuremath{\mathcal E}}}

\graphicspath{{./Figures/}}

\let\vec\bm

\begin{document}

\title{Effective electric field: quantifying the sensitivity of searches for new P,T-odd physics with EuCl$_3\cdot 6$H$_2$O.}

\author{A.~O.~Sushkov}\email{asu@bu.edu}
\affiliation{Department of Physics, Boston University, Boston, MA 02215, USA}
\affiliation{Department of Electrical and Computer Engineering, Boston University, Boston, MA 02215, USA}
\affiliation{Photonics Center, Boston University, Boston, MA 02215, USA}
\author{O.~P.~Sushkov}
\affiliation{School of Physics, UNSW, Sydney, Australia}
\author{A.~Yaresko}
\affiliation{Max-Planck-Institut fur Festkorperforschung, Heisenbergstrasse 1, D-70569 Stuttgart, Germany}

\begin{abstract}
Laboratory-scale precision experiments are a promising approach to searching for physics beyond the standard model. Non-centrosymmetric solids offer favorable statistical sensitivity for efforts that search for new fields, whose interactions violate the discrete parity and time-reversal symmetries.
One example is the electric Cosmic Axion Spin Precession Experiment (CASPEr-e), which is sensitive to the defining interaction of the QCD axion dark matter with gluons in atomic nuclei. The effective electric field is the parameter that quantifies the sensitivity of such experiments to new physics. We describe the theoretical approach to calculating the effective electric field for non-centrosymmetric sites in ionic insulating solids. We consider the specific example of the \chem{EuCl_3\cdot 6H_2O} crystal, which is a particularly promising material.
The optimistic estimate of the effective electric field for the $^{153}$Eu isotope in this crystal is $10\uu{MV/cm}$. The calculation uncertainty is estimated to be two orders of magnitude, dominated by the evaluation of the Europium nuclear Schiff moment. 
\end{abstract}

\maketitle

% Text spacing.
\baselineskip16pt

%\tableofcontents

\section{Introduction}

Non-centrosymmetric solids are a promising platform for experiments that search for violations of parity (P) and time reversal (T) symmetries, due to physics beyond the Standard Model. One possible approach is to search for permanent electric dipole moments (EDMs)~\cite{Safronova2018,Cairncross2017,ACME2018}. 
The strongest constraints on (non-Cabibbo–Kobayashi–Maskawa) hadronic P,T-violation are placed by experimental bounds on the permanent EDM of the neutron $d_n$~\cite{Abel2020}. 
The suggestion to use ferroelectrics to search for the permanent EDM dates back to Leggett~\cite{Leggett1978}. Since then, a number of solid material-based experiments have been developed, including efforts that use multiferroic materials~\cite{Sushkov2009,Sushkov2010,Rushchanskii2010,Eckel2012}, and atoms and polar molecules trapped in cryogenic noble gas matrices~\cite{Upadhyay2020,Li2022}. In this work our primary motivation is the CASPEr-electric search for the EDM and the gradient interactions of axion-like dark matter~\cite{Budker2014,Aybas2021a,Aybas2021b}. We consider a non-centrosymmetric crystal that contains no unpaired electron spins, but does contain an atomic species with non-vanishing nuclear spin $\v{I}$. Fundamental P,T-odd interactions can give rise to an energy shift $\delta\sE$ for such a nuclear spin, that depends on its orientation relative to a particular crystallographic direction~\cite{Mukhamedjanov2005a}. 
Since we focus on P,T-violation in nuclei, it is convenient to use the neutron EDM as the benchmark that quantifies this energy shift:
\begin{align}
\label{calE1}
\delta \sE= -d_n\v{E}^*\cdot\v{I}/I,
\end{align}
where $E^*$ is the effective electric field, the key parameter that determines the sensitivity of a given material to P,T-odd physics~\cite{Ludlow2013,Skripnikov2016,Flambaum2020,Aybas2021a}.
We note that $E^*$ is not a real electric field, in the sense that it is not sourced by electric charges and does not obey Maxwell's equations. It does have the same dimensions and the same discrete transformation properties as an electric field; the details of its physical origin are elucidated in this work.

There are two necessary conditions to have a nonzero $E^*$.
(i)~The discrete P,T symmetries are violated at the fundamenal level, in this work we consider the effect arising from the nuclear Schiff moment. (ii)~The crystalline site that hosts the nuclear spin $\v{I}$ is non-centrosymmetric. The P,T-odd effect that is described by Eq.~\eqref{calE1} is that it is energetically favorable for the nuclear spin $\v{I}$ to orient along the crystallographic direction, given by the orientation of the vector $\v{E}^*$.
In this work we calculate this energy shift for the Eu nuclear spins in \chem{EuCl_3\cdot 6H_2O}. This compound is under study as a promising candidate for the CASPEr-e experiment. The detailed experimental proposal is the subject of a separate manuscript~\bibnote[MIP]{Manuscript in preparation}. Here we briefly summarize the key features of this material that make it especially attractive.
The Eu $^7F_0\,\leftrightarrow\, ^5D_0$ optical transition at $579.7\uu{nm}$ wavelength is remarkably sharp: $25\uu{MHz}$ inhomogeneous linewidth has been observed in the stoichiometric crystal, isotopically purified in $^{35}$Cl~\cite{Ahlefeldt2016}.
Since this is smaller than the Eu hyperfine sublevel splittings, it may be possible to optically hyperpolarize the Eu nuclear spins in the entire crystal. There are two stable Eu isotopes, both with nuclear spin $I=5/2$: $^{153}$Eu has 52\% natural abundance and $^{151}$Eu has 48\% natural abundance. It is possible that the nuclear Schiff moment of $^{153}$Eu is strongly enhanced,
due to the closely-spaced opposite-parity nuclear energy levels, split by $\approx 100\uu{keV}$~\cite{Firestone1999}.
A low energy collective octupole mode 3$^-$ or even a static octupole deformation can further enhance the Schiff moment~\cite{Flambaum2020b,Dalton2023}.
We estimate that the $^{153}$Eu nuclear Schiff moment can be enhanced by a factor between $\approx5$ and $\approx100$, compared to that of $^{207}$Pb, which was used in the first-generation CASPEr-e experiments. This enhancement, along with the possibility of achieving a high degree of nuclear spin hyperpolarization, make \chem{EuCl_3\cdot 6H_2O} a promising candidate for future generations of CASPEr-e. 

In order to interpret future experimental measurements with \chem{EuCl_3\cdot 6H_2O} it is necessary to calculate the effective electric field $E^*$ in this compound. 
Our calculation is divided into three sections. In section \ref{sec:Schiff} we consider the nuclear Schiff moment of the two Eu stable isotopes and how it can arise from fundamental physics violating discrete P,T symmetries. In section \ref{sec:EuAtom} we describe the interaction between the nuclear Schiff moment and the Eu atomic electrons. In section \ref{sec:EuCrystal} we consider the Eu ion in the \chem{EuCl_3\cdot 6H_2O} crystal lattice and outline the solid-state calculation of its nuclear spin energy shift, extracting the magnitude and direction of the effective electric field.

\section{The Schiff moment of a nucleus}\label{sec:Schiff}

\subsection{Background}

\noindent
New hadronic parity (P) and time reversal (T) violating (P,T-odd) physics, such as the QCD theta parameter $\theta$, gives rise to P,T-odd nuclear interactions. 
The nucleon-nucleon P,T-odd interaction is usually parametrized in terms of pion exchange.
The pion-nucleon interaction vertex is
\begin{align}
	H_{\pi NN} &=g \bm{\pi}(\bar{N}\bm{\tau}i\gamma_5N)+\nonumber\\
	&+ {\bar g}_0\bm{\pi}(\bar{N}\bm{\tau}N)+{\bar g}_1\pi_0(\bar{N}N)+{\bar g}_2[\bm{\pi}(\bar{N}\bm{\tau}N)-3\pi_0(\bar{N}\tau_3N)],
	\label{eq:PTodd10}
\end{align}
where ${\bm \tau}$ is the isospin, ${\bm \pi}$ is the pion wavefunction, $N$ is the nucleon wavefunction, and $\gamma_5$ is the Dirac matrix. The first line in Eq.~(\ref{eq:PTodd10}) represents the usual strong interaction
which conserves the isospin, $g=14.1$~\cite{Ericson2002}. The second line represents the P,T-odd interaction.
%that does not conserve the isospin. 
In QCD the constants ${\bar g}_i$ ($i=0,1,2$) can be expressed in terms of the theta parameter~\cite{Yamanaka2017}, 
\begin{align}
\label{g012}
{\bar g}_0&=15.5 \times 10^{-3}\theta,\nonumber\\
{\bar g}_1&=-3.4 \times 10^{-3}\theta,\nonumber\\
{\bar g}_2&\approx 0.
\end{align}
The interaction (\ref{eq:PTodd10}) gives rise to the permanent neutron electric dipole moment (EDM)~\cite{Pospelov1999b,Yamanaka2017}.
\begin{align}
	d_n = -2.7\times 10^{-16}\,\theta \uu{e\cdot cm},
	\label{eq:energydn}
\end{align}
The interaction in Eq.~(\ref{eq:PTodd10}) also results in the P,T-odd nucleon-nucleon interaction. 
In a heavy nucleus with a single valence nucleon, within the non-relativistic approximation, we can average  
over the frozen core and be left with the P,T-odd effective single-particle potential for the valence nucleon:
\begin{align}
	W = \frac{G_F}{\sqrt{2}}\frac{\eta_a}{2m}\bm{\sigma}\cdot\nabla \rho_N(\bm{x}),
	\label{eq:PTodd40}
\end{align}  
where $a=n$ (neutron) or $a=p$ (proton), $m$ is the nucleon mass, $\bm{\sigma}$ is its spin, and $\rho_N(\bm{x})$ is the density of core nucleons~\cite{Sushkov1984}. The effective dimensionless coupling constant is given by
\begin{align}
	\eta_n = -\eta_p = 0.7\frac{\sqrt{2}}{G_Fm_{\pi}^2} g \left( -\frac{N-Z}{N+Z}\bar{g}_0 + \bar{g}_1 + 2\frac{N-Z}{N+Z}\bar{g}_2\right),
	\label{eq:Schiff45}
\end{align}
where $N$ is the number of neutrons and $Z$ is the number of protons in the nucleus; for many nuclei $(N-Z)/(N+Z)\approx 0.2$. The approximate factor $\sim 0.7$ in
Eq.~(\ref{eq:Schiff45}) arises from numerical averaging of the pion exchange over the shell model wave functions~\cite{Sushkov1984,Dmitriev1994}.

The potential (\ref{eq:PTodd40}) allows us to calculate the EDM of a nucleus:
\begin{align}
	\bm{d}_N = \langle 0|\bm{d}_N|0\rangle = 2\sum_n \frac{\langle 0|W|n\rangle\langle n|e\bm{x}|0\rangle}{E_0-E_n},
	\label{eq:PTodd50}
\end{align}
where $|0\rangle$ is the nuclear ground state of energy $E_0$, and the sum is over excited states $|n\rangle$ with energies $E_n$, having opposite parity compared to $|0\rangle$.

Does this nuclear EDM give rise to linear Stark shift for an atom in an applied external electric field $\bm{E}_0$? The answer is yes, but caution is advised. A naive expectation might be that the atomic Hamiltonian is modified: $H_{atom}\rightarrow H_{atom} - \bm{d}_N\cdot\bm{E}_0$.
This is wrong. According to the Schiff theorem, there is no first-order Stark shift due to the nuclear EDM $\bm{d}_N$~\cite{Schiff1963}. Under the assumption of a point-like nucleus, the atomic electron wavefunctions are perturbed, such that, in the new atomic ground state, the electric field at the nucleus vanishes. The dominant non-vanishing effect is due to the finite size of the nucleus~\cite{Sushkov1984}. The effect is parametrized by the P,T-odd nuclear Schiff moment:
\begin{align}
	\bm{S} = S\bm{I}/I = \frac{1}{10}\left(\langle x^2\bm{x}\rangle - \frac{5}{3}\langle x^2\rangle_q\,\bm{d}_N\right),
	\label{eq:Schiff10}
\end{align}
where $\langle x^2\rangle_q$ is the nuclear mean squared  electric charge radius, and the other terms are calculated using the 
P,T-odd correction to the nuclear charge density $\delta\rho({\bm x})$, which is due to the P,T-odd interaction in Eq.~\eqref{eq:PTodd40}.
Specifically:
%$\langle x^2\bm{x}\rangle$, $\langle\bm{d}\rangle$ are expectation values
\begin{align}
	\langle x^2{\bm x}\rangle = \int x^2{\bm x}\delta\rho({\bm x})d^3x, \ \ \ 
 	{\bm d}_N = \int \bm{x}\delta\rho({\bm x})d^3x, 
	\label{eq:Schiff20}
\end{align}
where ${\bm d}_N$ is the nuclear EDM, given by Eq.~\eqref{eq:PTodd50}.
%that are due to the  P,T-odd correction to the nuclear charge density $\delta\rho_{PT}({\bm x})$, which is due to the P,T-odd interaction in Eq.~\eqref{eq:PTodd40}.
The second term in Eq.~(\ref{eq:Schiff10}) originates from the Schiff screening by atomic electrons~\cite{Sushkov1984}. 

The effect of the Schiff moment on the atomic electrons is described by the P,T-odd electrostatic potential
of the nucleus
\begin{align}
	V(\bm{r}) = 4\pi(\bm{S}\cdot\nabla)\delta(\bm{r}),
	\label{eq:Schiff30}
\end{align}
where $\bm{r}$ is the electron coordinate~\cite{Sushkov1984}.
We note that the definition of the Schiff moment in Ref.~\cite{Khriplovich1997} differs from this one by a factor of $4\pi$.

As a naive order-of-magnitude estimate, one might expect $S\approx a^2_N d_N$, where $a_N$ is the nuclear radius, and the atomic energy shift on the order of $\bm{S}\cdot{\bm E}_0/a_0^2$, where $a_0$ is the Bohr radius and $E_0$ is the external electric field.
%{\cred  and $E_0$ effective electric field due to the noncenrocymmetric position of the ion in the lattice.}
If this were true, the energy shift due to the Schiff moment would be suppressed by a small factor of order $a_N^2/a_0^2\approx 10^{-8}$. Fortunately, due to the enhancement of the relativistic electron wavefunction at the nucleus, this suppression is offset by a factor $\approx \mathcal{R}Z^2\approx 10^5$, where $\mathcal{R}$ is the relativistic factor, see Eq.~\eqref{eq:matrixelement}. Thus the estimate for the energy shift of a neutral atom in an external electric field $E_0$ is $(\mathcal{R}Z^2/a_0^2) {\bm S}\cdot{\bm E}_0$.

\subsection{Estimates of the Schiff moments of $^{153}$Eu and $^{151}$Eu}

\noindent
The Schiff moment arises due to the mixing between nuclear quantum states of opposite parity.
The $^{153}$Eu nucleus is deformed, because there are clear rotational towers in its excitation spectrum~\cite{Firestone1999}. The standard theoretical description of this nucleus is based on the Nilsson model of a quadrupolar-deformed nucleus. In agreement with experimental data, the model predicts
the spin and parity of the ground state, $5/2^+$. It also
predicts the existence of the low-energy excited state with opposite parity, $5/2^-$. The wave functions of the odd proton in the Nilsson scheme are $|5/2^+\rangle=|413\frac{5}{2}\rangle$,
$|5/2^-\rangle=|532\frac{5}{2}\rangle$.
The experimentally-measured energy splitting between these sates is $97.4\uu{keV}$~\cite{Firestone1999}.

The $^{151}$Eu nucleus does not manifest clear rotational spectra at low angular momenta. However, the ground state
is still $|5/2^+\rangle$. Therefore, it is reasonable to assume that the Nilsson model is still relevant. The opposite parity state in this nucleus is at higher energy: $E_{5/2^-}=350\uu{keV}$.

The generic estimate for the Schiff moment of a heavy spherical nucleus is $S \approx 10^{-8}\eta\uu{e\,fm^3}$~\cite{Sushkov1984}.
The energy splitting  between opposite parity states in $^{153}$Eu,
$\Delta E=97.4\uu{keV}$, is 100 times smaller than that in spherical nuclei where $\Delta E \approx 8\uu{MeV}$.
Therefore, naively, one can expect that due to the small energy denominator  in $^{153}$Eu the
Schiff moment is enhanced by two orders of magnitude.
However, the overlap between the Nilsson single particle states $|5/2^+\rangle=|413\frac{5}{2}\rangle$ and $|5/2^-\rangle=|532\frac{5}{2}\rangle$
is small, $\approx 10^{-2}$~\cite{Bohr1998}. As a result, in spite of the
small energy denominator, in the single particle picture, the Schiff moment of $^{153}$Eu is practically the same as that of a spherical nucleus~\cite{Sushkov1984}.
However, there can be a collective enhancement by an order of magnitude~\cite{Sushkov1984}.
Hence, we arrive to the estimate that we call conservative:
\begin{align}
	S_c(^{153}\rm{Eu})\sim 10\times 10^{-8}\eta\uu{e\,fm^3}\approx 10^{-7}\eta\uu{e\,fm^3}.
	\label{eq:Schiff42}
\end{align}
The conservative estimate for the Schiff moment of the $^{151}$Eu nucleus is a factor of 3 smaller, due to the correspondingly larger energy denominator $\Delta E$.

There is an alternative description of the structure of low energy quantum states of the $^{153}$Eu nucleus~\cite{Flambaum2020b,Dalton2023}. The view is not based on the Nilsson  scheme.
Within this alternative approach the nucleus has a pear-shape static octupole deformation.
The single-particle proton state $|\frac{5}{2}^+\rangle$ and the opposite-parity state $|\frac{5}{2}^-\rangle$
are the {\it same} single-particle state, with the only difference being the global rotation of the pear.
It is unlikely that the Nilsson model is completely invalid for this nucleus, because it does correctly predict the ground state quantum numbers $|5/2^+\rangle$, which is the Nilsson state $|413\frac{5}{2}\rangle$. However, for some reason, the Nilsson state $|5/2^-\rangle=|532\frac{5}{2}\rangle$ does not exist in this approach. If we accept these assumptions, the overlap of single-particle components of the $|\frac{5}{2}^+\rangle$ and the $|\frac{5}{2}^-\rangle$
states is 100\%, which leads to the dramatic enhancement of $^{153}$Eu Schiff moment~\cite{Flambaum2020b}:
\begin{align}
	S_o(^{153}\rm{Eu})\approx 10^{-5}\eta\uu{e\,fm^3}.
	\label{eq:Schiff40}
\end{align}
This corresponds to $\times10^3$ enhancement, compared to the typical $\approx 10^{-8}\eta\uu{e\,fm^3}$ Schiff moment of a spherical nucleus, such as $^{199}$Hg and $^{129}$Xe. This is likely the most optimistic possible value for the $^{153}$Eu Schiff moment.
As before, the optimistic estimate for $^{153}$Eu is a factor of 3 lower than Eq.~(\ref{eq:Schiff40}).

The optimistic estimate in Eq.~(\ref{eq:Schiff40}) is two orders of magnitude larger than the conservative estimate in Eq.~(\ref{eq:Schiff42}). The true answer is likely to be somewhere between these two estimates, but more accurate calculations are needed to reduce the uncertainty. A reliable first-principles calculation is likely impossible, but a phenomenological approach, based on a fit of experimentally-measured E1-transition amplitudes, could work.
In any case, the conservative estimate (\ref{eq:Schiff42})
is at least a factor of 5 greater than the $^{207}$Pb Schiff moment estimate in Ref.~\cite{Aybas2021a}.

Next we need to express the Schiff moment in terms of the QCD $\theta$-parameter. Using Eqs.~(\ref{eq:Schiff45}) and (\ref{g012})
\begin{align}
\eta = \eta_p=0.5\times10^6\theta
	\label{eq:etatheta}
\end{align}
This value has to be used with Eqs.~(\ref{eq:Schiff40}) and (\ref{eq:Schiff42}).

\section{Interaction of the nuclear Schiff moment with electrons in an isolated Eu$^{3+}$ ion}\label{sec:EuAtom}

\subsection{The calculation of electron wavefunctions at the nucleus of an isolated Eu$^{3+}$ ion}

\noindent
In the following section~\ref{sec:EuCrystal} we will calculate the energy shift of a Eu nuclear spin in \chem{EuCl_3\cdot 6H_2O}, due to the nuclear Schiff moment. In this calculation we will make use of the Eu atom $6s$ and $6p$ electron wavefunctions near the nucleus.
In the current section we present an approximate treatment for the non-relativistic wavefunctions, that extends the
analysis presented in Ref.~\cite{Khriplovich1997}.

Let us define the effective principal quantum number $\nu$ that determines the outer electron energy:
\begin{align}
	\epsilon = -\frac{Z_i^2}{\nu^2}\frac{e^2}{2a_0},
	\label{eq:nu1}
\end{align}
where $e^2/2a_0=13.6\uu{eV}$ is the Rydberg energy, $a_0$ is the Bohr radius, and $Z_i=3$ is the ionic core charge of the Eu$^{3+}$. 
%(Q: why is $Z_i=1$ for as neutral atom? Seems like singly-charged ion is same as a neutral atom? Could $Z_i$ be absorbed into $\nu$?)\look 
In terms of the other quantum numbers
\begin{align}
	\nu = n_r+l+1-\sigma_l,
	\label{eq:nu2}
\end{align}
where $n_r$ is the radial quantum number, $l$ is the orbital angular momentum, and $\sigma_l$ is the quantum defect.

Let us consider the spatial region near the nucleus: $r\lesssim a_0/Z$, where $Z=63$ for Eu. In this region the nuclear Coulomb potential is unshielded and the radial non-relativistic electron wavefunctions can be approximated as:
\begin{align}
	\begin{split}
		R_{6s}(r\ll a_0/Z) &\approx A_s, \\
		R_{6p}(r\ll a_0/Z) &\approx A_p\frac{r}{a_0},
		\label{eq:wf1}
	\end{split}
\end{align}
where $A_s$ and $A_p$ are normalization constants.

Let us now consider the spatial region $a_0/Z\ll r\ll a_0/Z_i$. Here the WKB approximation holds (see Ref.~\cite{Budker2004}), and the radial wavefunction can be written in the semiclassical form:
\begin{align}
	R(r) = \frac{B}{r\sqrt{p}}\sin{\phi(r)},
	\label{eq:wf2}
\end{align}
where $B$ is a normalization constant, $\phi(r)$ is the semiclassical phase, and $p$ is the electron momentum given by:
\begin{align}
	p(r) = \sqrt{2m\left[\epsilon-V_a(r)-\frac{(l+1/2)^2}{2mr^2}\right]},
	\label{eq:wf3}
\end{align}
where $V_a$ is the self-consistent atomic potential. 
In order to find the constant $B$ we note that the wavefunction oscillates in the spatial region between the inner and the outer turning points $r_1$, $r_2$, and decays exponentially outside this region~\cite{Landau1981}.
%\bibnote{LandauIII, section 48.}. 
Therefore the wavefunction normalization integral is dominated by this spatial region 
\begin{align}
	1\approx\int_{r_1}^{r_2}R^2r^2\,dr \approx \frac{B^2}{2}\int_{r_1}^{r_2}\frac{dr}{p}.
	\label{eq:wf4}
\end{align}
To calculate this integral we write the Bohr quantization rule for radial motion:
\begin{align}
	\int_{r_1}^{r_2}p\,dr =\pi\hbar(n_r+\beta),
	\label{eq:wf5}
\end{align}
where $n_r$ is the radial quantum number and $\beta$ is the quantum defect.\look
Next we differentiate with respect to $n_r$. To take the derivative on the left-hand-side we use:
\begin{align}
	\frac{dp}{dn_r} = \frac{dp}{d\epsilon}\frac{d\epsilon}{dn_r} = \frac{m}{p}\frac{2Z_i^2}{\nu^3}\frac{e^2}{2a_0},
	\label{eq:wf6}
\end{align}
where in the last step we used eq.~\eqref{eq:nu1}. Substitution into eq.~\eqref{eq:wf5} gives:
\begin{align}
	\frac{2mZ_i^2}{\nu^3}\frac{e^2}{2a_0}\int_{r_1}^{r_2}\frac{dr}{p} =\pi\hbar.
	\label{eq:wf7}
\end{align}
Comparing with eq.~\eqref{eq:wf4} and using $e^2/2a_0=\hbar^2/(2ma_0^2)$, we get:
\begin{align}
	B = \frac{Z_i}{a_0}\sqrt{\frac{\hbar}{\pi\nu^3}}.
	\label{eq:wf8}
\end{align}

We match the wavefunctions \eqref{eq:wf1} and \eqref{eq:wf2} at $r\approx a_0/Z$. At this radius the momentum $p\approx \hbar/r\approx \hbar Z/a_0$. Dropping factors of order unity this gives:
\begin{align}
	\begin{split}
		A_s &= B\sqrt{\frac{Z}{\hbar a_0}} = Z_i\sqrt{\frac{Z}{a_0^3\nu_s^3}},\\
		A_p &= ZB\sqrt{\frac{Z}{\hbar a_0}} = Z_iZ\sqrt{\frac{Z}{a_0^3\nu_p^3}}.
		\label{eq:wf9}
	\end{split}
\end{align}

For completeness we write the full expressions for the non-relativistic radial wavefunctions near the origin, including numerical factors taken from Ref.~\cite{Khriplovich1997}:
\begin{align}
	\begin{split}
		R_{6s}(r\ll a_0/Z) &= 2Z_i\left(\frac{Z}{a_0^3\nu_{6s}^3}\right)^{1/2},\\
		R_{6p}(r\ll a_0/Z) &= \frac{2}{3}Z_iZ\left(\frac{Z}{a_0^3\nu_{6p}^3}\right)^{1/2}\frac{r}{a_0}.
		\label{eq:wf10}
	\end{split}
\end{align}

The values of the effective principal quantum numbers $\nu$ for the Eu$^{3+}$ 6s and 6p electrons can be extracted from their ionization energies
\bibnote{From the NIST atomic spectra database for Eu III ion we extract the energy of the 4f$^6$6s state to be $49 000\uu{cm^{-1}}$ and the 4f$^6$6p$_{1/2}$ state to be $82 000\uu{cm^{-1}}$.
Then we subtract these from the ionization limit of $200 000\uu{cm^{-1}}$.
}:
\begin{align}
	\epsilon_{6s} &=-151000\uu{cm^{-1}}\,\ra\,\nu_{6s}=2.60,\nonumber\\
	\epsilon_{6p_{1/2}} &=-118000\uu{cm^{-1}}\,\ra\,\nu_{6p_{1/2}}=2.92.	
	\label{eq:nu2}
\end{align}

\subsection{The matrix element of the Schiff moment potential for an isolated Eu$^{3+}$ ion with a single electron}

\noindent
The nuclear Schiff moment $\vec{S} = S\,\vec{I}/I$, where $I$ is the nuclear spin, creates the electrostatic potential 
$V(\vec{r})$, see Eq.~(\ref{eq:Schiff30}). This potential creates the following perturbation acting on an electron
\begin{align}
	\delta H(\vec{r}) = -|e|V(\vec{r})=-|e| 4\pi(\vec{S}\cdot\vec{\nabla})\delta(\vec{r}),
	\label{eq:VSchiff}
\end{align}
where $-|e|$ is the electron charge.
The Schiff moment couples to the gradient of the electron wavefunction at the nucleus.

The electron configuration of Eu$^{3+}$ is [Xe]$4f^6$, and the ground state is $^7\hspace{-0.1em}F_0$.
Let us outline the calculation of the matrix element $\bra{6s}V(\vec{r})\ket{6p_z}$ for an isolated Eu$^{3+}$ ion. We will use this matrix element in the following section~\ref{sec:EuCrystal}. We note that the choice of the $6s$ and $6p$ wavefunctions is somewhat arbitrary, and we could have chosen $s$ and $p$ wavefunctions with any principal quantum number $\geq 6$.

Due to the delta-function in Eq.~(\ref{eq:VSchiff}), we can use the wavefunctions near the nucleus, given by Eq.~\eqref{eq:wf10}. In addition to the radial wavefunctions, we need the spherical harmonics: $Y_{00}=\sqrt{1/4\pi}$ and $Y_{10}=\sqrt{3/4\pi}\cos{\theta}$. The matrix element then reduces to the integral
\begin{align}
	\bra{6s}V\ket{6p_z} &= 4\pi S_z \frac{4}{3}\frac{Z^2Z_i^2}{a_0^3(\nu_{6s}\nu_{6p})^{3/2}}
	\int\frac{r}{a_0}\sqrt{\frac{1}{4\pi}}\sqrt{\frac{3}{4\pi}}\cos{\theta}\left[\frac{\partial}{\partial z}\delta(\vec{r})\right]\,d^3r \nonumber\\
	&= S_z \frac{4}{\sqrt{3}}\frac{Z^2Z_i^2}{a_0^4(\nu_{6s}\nu_{6p})^{3/2}}\int\left[\frac{\partial}{\partial z}\delta(\vec{r})\right]z\,d^3r,
	\label{eq:int1}
\end{align}
where we used $r\cos{\theta}=z$.
Integrating by parts:
\begin{align}
	\int\left[\frac{\partial}{\partial z}\delta(\vec{r})\right]z\,d^3r =
	-\int\delta(\vec{r})\,d^3r = -1.
	\label{eq:int2}
\end{align}

The full expression for the atomic matrix element is:
\begin{align}
	\bra{6s}V\ket{6p_z} = -\frac{4}{\sqrt{3}} S_z \frac{Z^2Z_i^2}{a_0^4(\nu_{6s}\nu_{6p})^{3/2}}\sR,
	\label{eq:matrixelement}
\end{align}
where $\sR$ is the relativistic factor that arises when calculating with the full Dirac relativistic wavefunction~\cite{Sushkov1984}. In the non-relativistic limit $Z\alpha\ra 0$, $\sR\ra 1$. 
Because we consider the electrons with definite values of the $l,s$ quantum numbers, we use Clebsch-Gordan coefficients to express the relativistic factor as a linear combination of factors for electrons with definite value of the total angular momentum $j$:
\begin{align}
	\sR = \frac{1}{3}\sR_{1/2}+\frac{2}{3}\sR_{3/2},
	\label{eq:R1}
\end{align}
where $\sR_{1/2}$ and $\sR_{3/2}$ are the relativistic factors for the $p_{1/2}$ and $p_{3/2}$ electrons respectively. In turn, these are given in Ref.~\cite{Sushkov1984}:
\begin{align}
	\sR_{1/2} &\approx \frac{4\gamma_{1/2}x_0^{2\gamma_{1/2}-2}}{[\Gamma(2\gamma_{1/2}+1)]^2},
 \nonumber \\
	\sR_{3/2} &\approx \frac{48\gamma_{1/2}x_0^{\gamma_{1/2}+\gamma_{3/2}-3}}{\Gamma(2\gamma_{1/2}+1)\Gamma(2\gamma_{3/2}+1)}.	
	\label{eq:R2}
\end{align}
Here $\gamma_{1/2} = \sqrt{1-Z^2\alpha^2}$, $\gamma_{3/2} = \sqrt{4-Z^2\alpha^2}$, and $x_0=2Za_N/a_0$, where $a_N$ is the nuclear radius.

We have calculated the atomic matrix element of the Schiff moment interaction.
In order to evaluate the energy shift in a \chem{EuCl_3\cdot 6H_2O} crystal, we need to consider how the crystal electronic wavefunctions behave near the Eu nuclei. We will do this by expanding them in terms of the Eu atomic wavefunctions.

\section{A \chem{Eu^{3+}} ion in the \chem{EuCl_3\cdot 6H_2O} crystal}
\label{sec:EuCrystal}

\subsection{Crystal structure of \chem{EuCl_3\cdot 6H_2O}}\label{sec:33}

\begin{figure}[h!]
	\centering
	\includegraphics[width=0.4\textwidth]{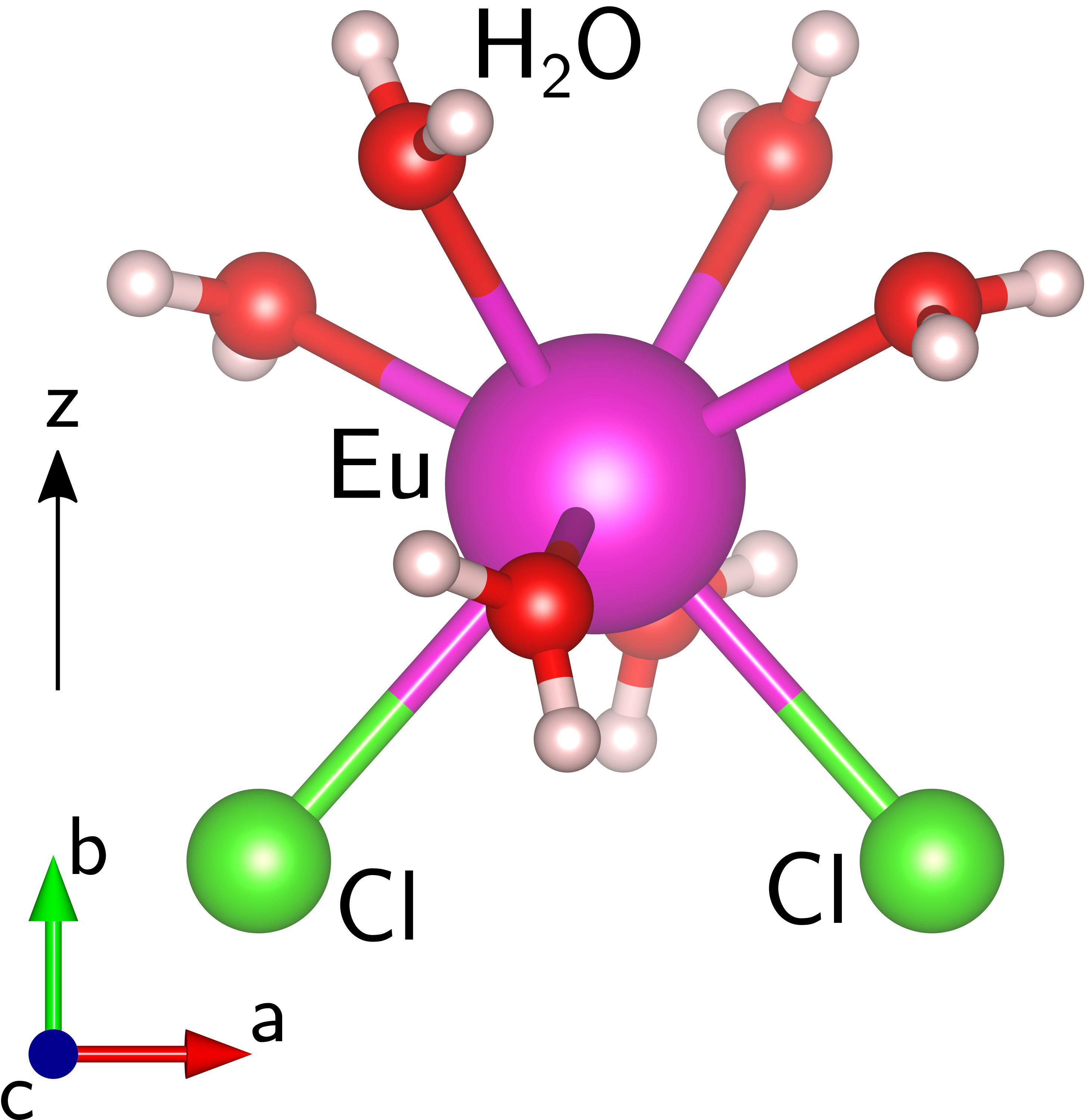}
	\caption{\fontsize{10}{11}\selectfont
		The structure of the cationic \chem{[Eu(H_2O)_6Cl_2]^+} unit in \chem{EuCl_3\cdot 6H_2O}~\cite{Tambornino2014}.}
	\label{fig:200}
\end{figure}

The crystal structure of \chem{EuCl_3\cdot 6H_2O} at 293~K is monoclinic, space group P2/n~\cite{Tambornino2014}. The molar mass is 366.41. The Eu$^{3+}$ ion sites have C2 symmetry, with the axis corresponding to the b-axis of the crystal. The corresponding lattice constant is $b=6.5322\uu{\AA}$. The unit cell has two Eu sites, that are symmetric conjugates of each other.
The local environment of each Eu ion is highly asymmetric, Fig.~\ref{fig:200}. This suggests that there may be a substantial effective electric field.
We choose the coordinate system with the origin at the Eu site, and the z-axis along the C2 symmetry axis, as shown in Fig.~\ref{fig:200}. The two nearest Cl ions are at negative z. The third Cl ion is further away and is not shown.

\subsection{Electron wavefunctions in the \chem{EuCl_3\cdot 6H_2O} crystal}\label{sec:34}

\begin{figure}[h!]
	\centering
	\includegraphics[width=0.8\textwidth]{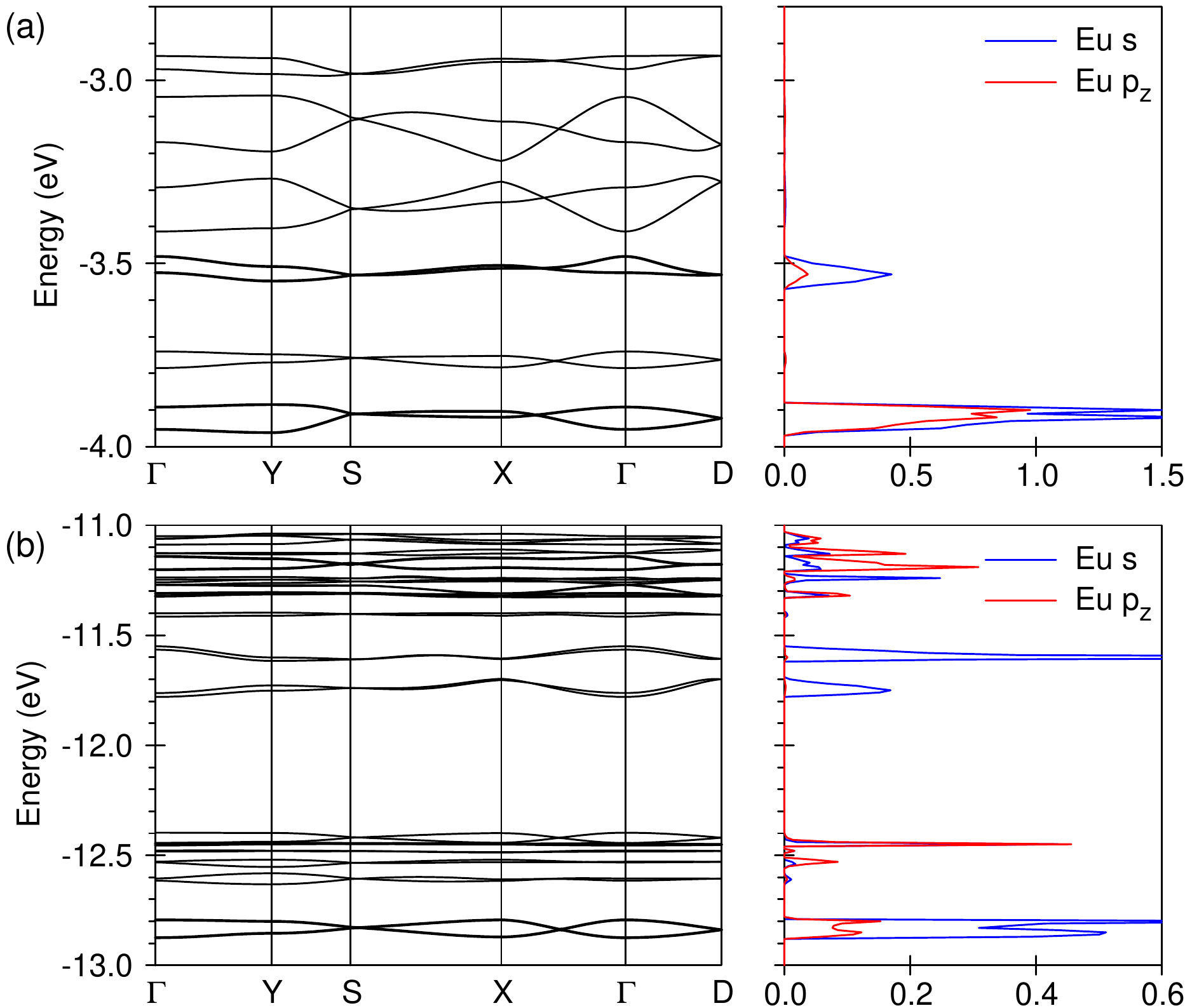}
	\caption{\fontsize{10}{11}\selectfont
		Electronic band structure of \chem{EuCl_3\cdot 6H_2O}, the energy is presented with respect to the chemical potential.
		(a) Bands formed by $p$-electrons of Cl.
		(b) Bands formed by $p$-electrons of Ne (H$_2$O). 
		The right sub-panel in each panel gives a spectral weight of the corresponding band when it is decomposed in terms of $s$ and $p_z$ orbitals
		of the Eu$^{3+}$ ion.
	}
	\label{fig:400}
\end{figure}

\noindent
Our goal is to calculate the Eu nucleus energy shift due to the interaction of crystal electrons with the nuclear Schiff moment. To do so we need the
many-body electron wave function of the crystal. There are two methods to approach this problem. (i) The finite cluster method, used previously for  calculation of the similar effect for Pb nucleus in ferroelectric \chem{PbTiO_3} or PMN-PT crystal~\cite{Ludlow2013,Skripnikov2016}.
(ii) The band structure method, used in the present work.
For the band structure calculations we use the Linear Muffin-Tin Orbital (LMTO) method. The code is described in Ref.~\cite{Antonov2004}.
Since it is difficult to treat a water molecule with very short
oxygen-proton distance by the LMTO code, we replace water molecules by Ne atoms  
which have the same electronic configuration as H$_2$O.
The band structure of \chem{EuCl_3\cdot 6H_2O} calculated in this way is presented in Fig.~\ref{fig:400}.
The energy difference between the Cl $p$-bands and the
Ne $p$-bands is $\approx12-3.5=8.5\uu{eV}$. In order to verify our approximation, we compare this with the difference between the water molecule ionization energy and the electron affinity in the Cl$^-$ ion: $12.6-3.6=9\uu{eV}$. These values are close, which means that, for calculation of the electron band energies, the H$_2$O molecule can be replaced with the Ne atom.

Naively one might expect that, having calculated the crystal electronic wavefunctions and band energies, it should be possible to directly calculate the expectation value of the Schiff interaction in Eq.~(\ref{eq:VSchiff}), giving the energy shift we are after. The problem is that the dominant contribution to this energy shift is from the spatial region inside the 
Eu nucleus, where the electron dynamics are ultra-relativistic.
No existing band structure calculation can provide accurate wave functions down to such small distances from the atomic nucleus.
To overcome this problem we match the band structure wave functions with wave functions of an isolated Eu ion.
%at distances $r< 0.7\AA$. 
As soon as matching coefficients are established we use the analytic results, obtained
for states of an isolated ion, section~\ref{sec:EuAtom}. As matching wave functions we use $6s$ and $6p_z$ states
calculated in a frozen Eu$^{3+}$ core. The states with higher orbital angular momenta (d,f,...) are also present, however they do not contribute to the Schiff moment interaction because they do not penetrate into the nucleus.

\begin{figure}[t!]
	\centering
	\includegraphics[width=0.48\textwidth]{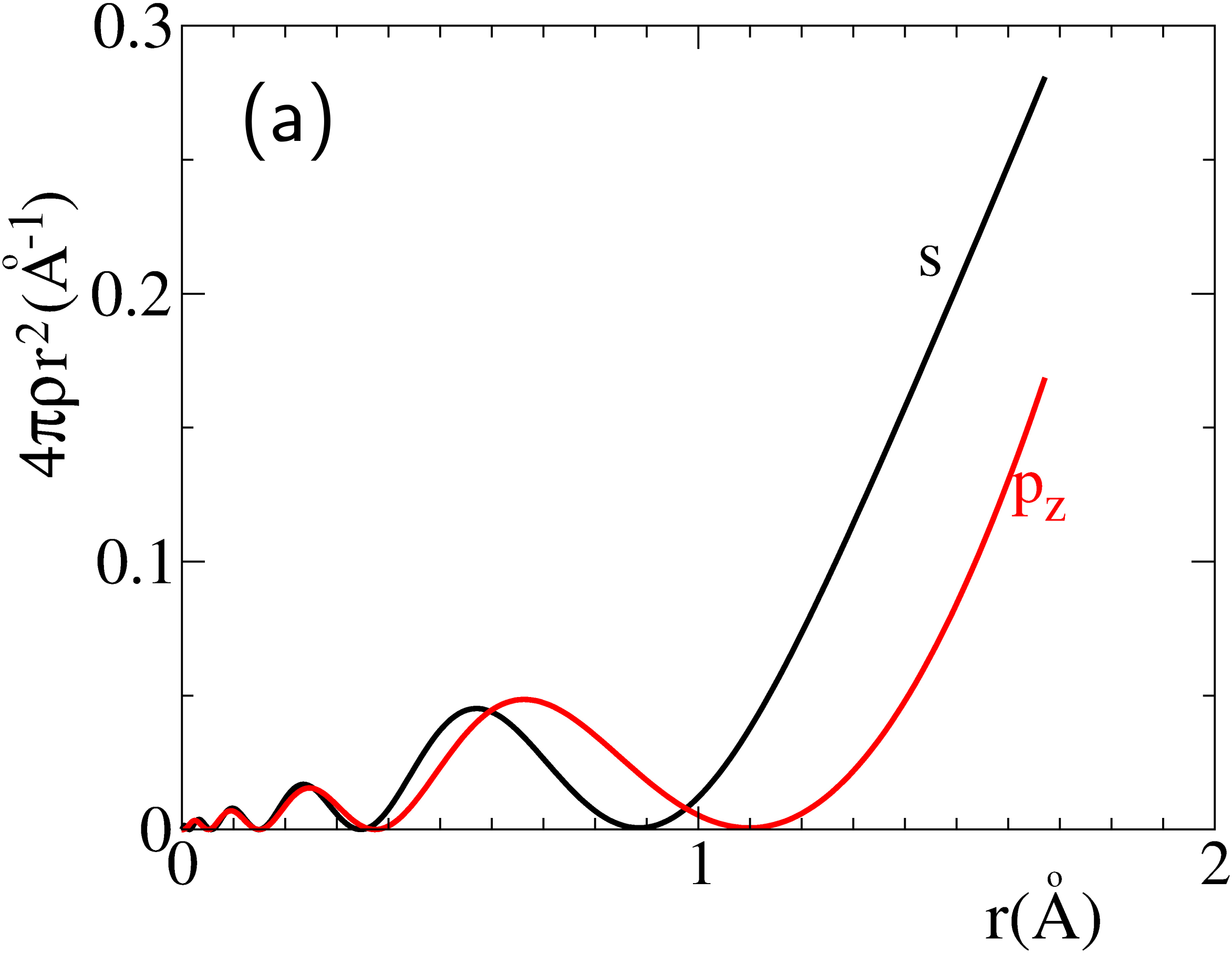}
	\includegraphics[width=0.48\textwidth]{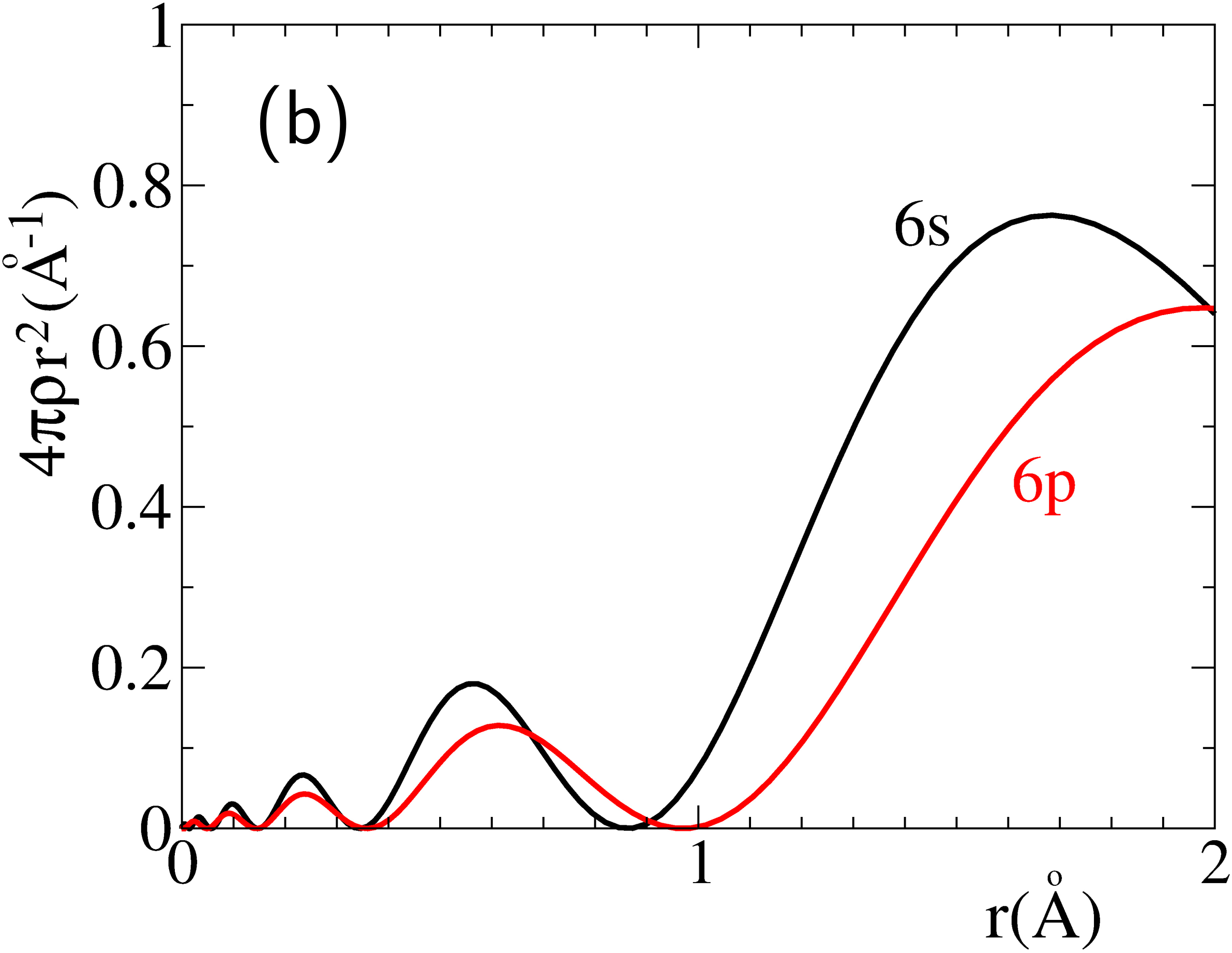}
	\caption{\fontsize{10}{11}\selectfont
		(a) Band structure projected electron densities near Eu ion. The densities are defined in Eq.~(\ref{spd}).
		(b) Electron densities corresponding to atomic 6s and 6p orbitals calculated in the frozen core of Eu$^{3+}$ ion.
	}
	\label{fig:300}
\end{figure}

We define the band structure $s$- and $p_z$-wave electron densities near the Eu ion as:
\begin{eqnarray}
\label{spd}
&&\rho_s(r)=\sum_k|\langle\psi_k({\bm r})|s\rangle|^2\nonumber\\
&&\rho_p(r)=\sum_k|\langle\psi_k({\bm r})|p_z\rangle|^2,
\end{eqnarray}
where $k$ is the quasi-momentum, $\psi_k$ are the band structure wave functions, and $|s\rangle$ and $|p_z\rangle$ are the spherical harmonics $Y_{00}$, $Y_{10}$, centered at the Eu site. The summation is performed over all filled bands, $E_k < 0$. Plots of these densities, resulting from our band structure calculations, are presented in Fig.~\ref{fig:300}(a).
In Fig.~\ref{fig:300}(b) we present the electron densities corresponding to atomic 6s and 6p wave 
functions, calculated in the approximation of the frozen Eu$^{3+}$ core with electronic configuration $1s^2...4f^6$. 
The core itself is obtained by the Hartree-Fock procedure  averaged over polarizations of the open $4f^6$ shell~\bibnote{We thank V.~Dzuba for calculating the atomic 6s and 6p electron densities.}. As expected, near the Eu site (at $r < 1\AA$) the corresponding densities in panels (a) and (b) of Fig.~\ref{fig:300} are of similar shape but different amplitudes.
Let us compare the densities at $r=0.6\AA$, where their values are near a local maximum. The ratio of the s-wave densities is $\approx 0.24$, and the ratio of the p-wave densities is $\approx 0.44$. We take square roots of these numbers to obtain the coefficients of the expansion of the effective band wavefunction at $r<1\AA$:
\begin{align}
%\psi = 0.492\psi_{6s} \pm 0.662\psi_{6p_z}
	\psi = 0.49\psi_{6s} \pm 0.66\psi_{6p_z},
	\label{eq:crwf01}
\end{align}
Note that, for now, the sign is undetermined.

Equation (\ref{eq:crwf01}) gives the total weights, but does not account for the many-body character of the crystal wave function.
The effect we consider arises from the interference of the Eu s- and p- waves. However the Eu is not an isolated ion, but is located in a crystal lattice, where states that have different quasi-momenta $k$ can not interfere, because they belong to different electrons, see Eq.~(\ref{spd}).
To address this crucial issue we must consider each sub-band separately.
We calculate the s- and p-spectral density for each sub-band, plotting the densities as a function of energy in the right-hand panels of Fig.~\ref{fig:400}. 
Let us enumerate the 14 sub-bands with the index $i\in\{1,...,14\}$, so that the wavefunction of each band near the Eu site can be represented as
\begin{align}
	\psi^{(i)} = a_s^{(i)}\psi_{6s}+a_p^{(i)}\psi_{6p_z},
	\label{eq:crwf05}
\end{align}  
where $a_s^{(i)},\,a_p^{(i)}$ are the expansion coefficients. 
For each sub-band the integrated spectral density $w^{(i)}_s\propto |a_s^{(i)}|^2$ and $w^{(i)}_p\propto |a_p^{(i)}|^2$ is proportional to the weight of this sub-band's contribution to the sum over quasi-momentum $k$ in Eq.~(\ref{spd}). 
The overall normalization of the contributions is determined by Eq.~\eqref{eq:crwf01}, so that 
\begin{align}
	|a_s^{(i)}| = 0.49\sqrt{\frac{w_s^{(i)}}{\sum_j w_s^{(j)}}},\,\,\,
	|a_p^{(i)}| = 0.66\sqrt{\frac{w_p^{(i)}}{\sum_j w_p^{(j)}}},
	\label{eq:crwf10}
\end{align}
where the index $j\in\{1,...,14\}$ also enumerates the bands.
The spectral weights $w^{(i)}_s$ and $w^{(i)}_p$, extracted for all 14 sub-bands from Fig.~\ref{fig:400}, are listed in Table~\ref{T1}.
\begin{table}[h!]
	\begin{center}
		\begin{tabular}{c|c|c|c|c|c}
			band index & band energy (eV) & $w_{s}$ & $a_s$ & $w_{p}$ & $a_p$ \\
			\hline 
			1 (Cl)&-3.5&0.02&0.166&0.004&0.145 \\
			2 (Cl)&-3.9&0.071&0.313&0.046&0.491 \\ 
			3 (Ne)&-11.0&0.0011&0.039&0.0021&-0.105 \\
			4 (Ne)&-11.05&0.0015&0.0455&0.0044&-0.152 \\
			5 (Ne)&-11.1&0.002&0.0526&0.009&-0.217 \\
			6 (Ne)&-11.25&0.0027&0.0611&0.0004&-0.046 \\
			7 (Ne)&-11.3&0.0008&0.0333&0.002&-0.102 \\
			8 (Ne)&-11.6&0.025&0.186&0&0 \\
			9 (Ne)&-11.7&0.008&0.105&0&0 \\
			10 (Ne)&-12.4&0.004&0.0744&0.0058&-0.174 \\
			11 (Ne)&-12.45&0.00012&0.0129&0.00017&-0.030 \\
			12 (Ne)&-12.55&0.0006&0.0288&0.0017&-0.0944 \\
			13 (Ne)&-12.6&0.00026&0.019&0.0001&-0.023 \\
			14 (Ne)&-12.8&0.038&0.229&0.008&-0.205 \\
			\hline 
			\multicolumn{2}{c|}{total area} & 0.175 & & 0.084 & \\
		\end{tabular}
		\caption{The values of the crystal band wave function expansion coefficients. The second column presents the energy of each of 14 bands shown in
	Fig.\ref{fig:400}. The third and the fifth columns give spectral weights $w_s$ and $w_p$
	of the bands, the overall spectral weight scale is arbitrary. The fourth and the sixth
	columns list the coefficients in the wave function (\ref{eq:crwf05}).
		}
		\label{T1}
	\end{center}
	\nonumber
\end{table}

Finally we have to determine the signs of the coefficients $a^{(i)}_{s,p}$. 
As shown in Fig.~\ref{fig:200}, the z axis of our coordinate system is along the crystal C2 axis, pointing in the direction from the nearest Cl ions towards the oxygen ions. 
Our atomic radial wave functions $R_{6s}(r)$ and $R_{6p}(r)$ are defined to be positive
as $r\to 0$, see Eq~\eqref{eq:wf10}. Therefore they have opposite signs at $r > 1\AA$, since the 6p wavefunction has an extra radial node. We choose the coefficients $a^{(i)}_{s}$ to be positive, their values are presented in Table~\ref{T1}. 
Let us consider the electron density z-asymmetry with respect to the Eu site, defined for each band $i$ as $\Delta n^{(i)} = |\psi^{(i)}(z>0)|^2-|\psi^{(i)}(z<0)|^2$, where $z$ is some typical interatomic distance, say $z\approx 1.5\AA$. We expand each $\psi^{(i)}$ into the 6s and the 6p wavefunctions, as in Eq.~\eqref{eq:crwf05}.
Since the electron densities of the 6s state $|\psi_{6s}|^2$ and of the 6p state $|\psi_{6p}|^2$ are spherically symmetric, only the cross-terms remain in the asymmetry:
$\Delta n^{(i)}\propto a_s^{(i)}a_p^{(i)}R_{6s}(r=|z|)R_{6p}(r=|z|)$.
The cross terms add instead of canceling, because the $\cos\theta$ factor in the spherical harmonic of the $p_z$-orbital is positive for $z>0$ and negative for $z<0$.
Consider the two Cl ions nearest to the Eu site, Fig.~\ref{fig:200}. Their electron density is shifted towards negative $z$, so for the Cl bands $\Delta n^{(i\in\{1,2\})}<0$. Keeping in mind that $a_s$ was defined to be positive and $R_{6s}(r)$ and $R_{6p}(r)$ have opposite signs, the coefficient $a_p^{(i)}$ is positive for the two Cl bands $i=1,2$.
Analogously, the Ne(H$_2$O) electron density is shifted towards positive $z$, so for the Ne bands $\Delta n^{(i\in\{3,...,14\})}>0$. Therefore the coefficient $a_p^{(i)}$ is negative for the twelve Ne bands $i=3,...,14$. These signs correspond to the values listed in Table~\ref{T1}.

\subsection{Calculation of the nuclear spin energy shift due to the Schiff moment}

\noindent
Having the multi-electron wave function determined in the previous subsection and taking the expectation value of the Schiff interaction
(\ref{eq:VSchiff}) we find the energy shift due to Eu Schiff moment
\begin{align}
	\delta\sE = -2|e|\sum_k \bra{\psi_k}V(\vec{r})\ket{\psi_k}
		 =-2|e|\sum_{i=1}^{14} \bra{\psi^{(i)}}V(\vec{r})\ket{\psi^{(i)}},
	\label{eq:energyCl1}
\end{align} 
where the factor of 2 is due to electron spin degeneracy.
We now use the expansion in Eq.~\eqref{eq:crwf05}:
\begin{align}
	\delta\sE = -4|e|\sum_{i=1}^{14}a_s^{(i)}a_p^{(i)}\bra{6s}V(\vec{r})\ket{6p_z}
	\label{eq:energyCl2}
\end{align} 
Finally using the coefficients from Table~\ref{T1} and the atomic matrix element given in Eq.~\eqref{eq:matrixelement} we find
\begin{align}
	\delta\sE&=-4|e|\times0.086\bra{6s}V(\vec{r})\ket{6p_z}\nonumber \\
	&=+4|e|\times0.086\times\frac{4}{\sqrt{3}} S_z \frac{Z^2Z_i^2}{a_0^4(\nu_{6s}\nu_{6p})^{3/2}}\left(\frac{1}{3}\sR_{1/2}+\frac{2}{3}\sR_{3/2}\right),
	\label{eq:energy1}
\end{align}
Converting to atomic units we get:
\begin{align}
	\frac{\delta\sE}{e^2/a_0} = +4\times0.086\times\frac{4}{\sqrt{3}} \frac{Z^2Z_i^2}{(\nu_{6s}\nu_{6p})^{3/2}}\left(\frac{1}{3}\sR_{1/2}+\frac{2}{3}\sR_{3/2}\right)\frac{S_z}{|e|a_0^3}.
	\label{eq:energy2}
\end{align}
We note that there is a degree of cancelation between the contributions from the Cl and the Ne (H$_2$O) electrons, see tab.~\ref{T1}. This warrants the careful calculation presented in our work.

\subsection{Calculation of the effective electric field in \chem{EuCl_3\cdot 6H_2O}}

The values of the parameters in Eq.~\eqref{eq:energy2} are given in Table~\ref{tab:4}.
\begin{table}[t!]
	\begin{center}
		\begin{tabular}{|c|c|c|c|c|c|}
			\hline
			$Z$ & $Z_i$ & $\nu_{6s}$ & $\nu_{6p}$ & $\sR_{1/2}$ & $\sR_{3/2}$ \\
			\hline 
			63 & 3 & 2.60 & 2.92 & 3.3 & 2.5 \\
			\hline 
		\end{tabular}
		\caption{The values of parameters in Eq.~\eqref{eq:energy2}.}
		\label{tab:4}
	\end{center}
\end{table} 
Substitution of these parameters into Eq.~\eqref{eq:energy2} gives the P,T-odd energy shift of Eu nucleus
\begin{eqnarray}
  \label{s2}
 {\delta \sE} = +1.1\times 10^5\frac{S_z}{|e|a_0^3}\uu{eV} 
\end{eqnarray}
This has to be compared with the result for the $^{207}$Pb nucleus in the PMN-PT crystal~\cite{Ludlow2013,Skripnikov2016},
$\delta\sE=-5.9\times 10^5\frac{S_z}{|e|a_0^3}\uu{eV}$. 
The numerical coefficients in these expressions have to be obtained from a quantum chemistry calculation, such as the one we describe in section~\ref{sec:EuCrystal}.
%The entire quantum chemistry is "hidden" in the coefficient 1.1 (5.9).
The difference in the absolute value of these coefficients is mainly due to the Z-scaling.
The difference in sign is due to the different choice of the positive direction of the z-axis.
In PMN-PT the positive direction is along the direction of increasing electron density.
For \chem{EuCl_3\cdot 6H_2O} we chose the z-axis pointing from the Eu to the O ions, along the crystal C$_2$ axis, Fig.~\ref{fig:200}. The electron density decreases along this direction.

%Of course Shiff moments of Eu nucleus and Pb nucleus are different, there is an enhancementin Eu as discussed in section ???. Besides that it is likely that \chem{EuCl_3\cdot 6H_2O} is much more efficient experimentally than PMN-PT due to the possibility of optical pumping~\cite{Sasha}.

Using Eq.~\eqref{s2} with the published optimistic estimate of the $^{153}$Eu nuclear Schiff moment (\ref{eq:Schiff40})
and together with the relation (\ref{eq:etatheta}) we arrive at the following energy shift of the $^{153}$Eu nuclear spin in \chem{EuCl_3\cdot 6H_2O}:
\begin{align}
%\label{sf1}
\delta\sE_o= 8\times 10^{-15}\eta=3\times 10^{-9}\theta\uu{[eV]} .
\label{eq:energy49}
\end{align}
Comparing this with Eq.~(\ref{calE1}) and Eq.~(\ref{eq:energydn}) we find the value of the effective electric field:
\begin{align}
E_o^*=10\uu{MV/cm}.
\label{eq:energy50}
\end{align}
The results (\ref{eq:energy49}), (\ref{eq:energy50}) are optimistic estimates for $^{153}$Eu.
Conservative estimates, based on the conservative $^{153}$Eu nuclear Schiff moment (\ref{eq:Schiff42}), are two orders of magnitude smaller.
We reiterate that the true answer is likely to be somewhere between these two limits. 
An accurate analysis of the nuclear part of the problem is needed to reduce the uncertainty.
For $^{151}$Eu the values are approximately 3 times smaller than for $^{153}$Eu.

\subsection{The two Eu sites}

The Eu site in \chem{EuCl_3\cdot 6H_2O} is noncentrosymmetric, so $E^*$ at each site is nonzero.
However, a unit cell contains two different Eu sites with opposite orientation.
In other words there are two different Eu sublattices with equal and opposite values of $E^*$.
Therefore, if we include both these sublattices and average the energy shift \eqref{eq:energy49} over the entire Eu ensemble in the crystal, the effect will vanish. In order to avoid this, we plan to apply an electric field to the crystal, to resolve the optical hyperfine transitions of the two different sublattices, and optically pump the $^{135}$Eu nuclear spins of only one of them. This is enabled by the remarkably narrow inhomogeneous linewidth of the Eu $^7F_0\,\rightarrow\, ^5D_0$ optical transition at $579.7\uu{nm}$ wavelength: linewidths of $25\uu{MHz}$ has been observed in the stoichiometric crystal, isotopically purified in $^{35}$Cl~\cite{Ahlefeldt2016}. The presence of the $^{151}$Eu isotope, with a different magnetic moment and different Schiff moment, also enables co-magnetometer measurements that control systematic effects. Experimental details are described in Ref.~\bibnote[MIP].

\section{Conclusion}
We consider the methodology for calculating the magnitude of nuclear P,T-odd effects in non-centrosymmetric crystalline solids containing heavy atomic species.
We focus on the crystal \chem{EuCl_3\cdot 6H_2O} as a promising candidate for 
CASPEr-electric experiment for searches of the electric dipole moment and the gradient interactions of axion-like dark matter.
The CASPEr-e search for axion dark matter will search for the spin precession of the $^{153}$Eu nuclear spin ensemble.
In the present work we calculate the magnitude of the effective electric field, which is necessary to calculate the magnitude of the expected signal. 
We address the possible enhancement of the $^{153}$Eu nuclear Schiff moment and perform the solid-state band structure calculation of the nuclear spin energy shift.
Our optimistic estimate shows a significant enhancement of the effective electric field, compared to, for example, $^{207}$Pb-containing ferroelectrics, which were used for first-generation CASPEr-e measurements. The uncertainty is dominated by the estimate of the $^{153}$Eu nuclear Schiff moment. 

\begin{acknowledgments}

\noindent
AOS acknowledges support by the National Science Foundation CAREER grant PHY-2145162, and the U.S. Department of Energy, Office of High Energy Physics program under the QuantISED program, FWP 100667.
OPS acknowledges support from the Australian Research Council Centre of Excellence in Future
Low-Energy Electronics Technology (FLEET) (Grant No.CE170100039).

\end{acknowledgments}

\bibliographystyle{apsrev4-2}
%\bibliography{library} 
%\end{document}


\end{document}